\documentclass[a4paper,11pt]{article}
\usepackage{pos}
\usepackage{bigints}

\newcommand{\MeV}{\, {\rm MeV}}
\newcommand \hmu {\hat{\mu}}
\newcommand{\calO}{{\cal O}}

\title{Equation of state of QCD at finite chemical potential from an alternative expansion scheme}

\author*[a,b]{Paolo Parotto}
\author[b]{Szabolcs Bors\'anyi}
\author[a,b,c,d]{Zoltan Fodor}
\author[b]{Jana N. Guenther}
\author[b]{Ruben Kara}
\author[e]{Sandor D. Katz}
\author[e]{Attila P\'asztor}
\author[f]{Claudia Ratti}
\author[b,d]{Kalman K. Szab\'o}

\affiliation[a]{Pennsylvania State University, Department of Physics, State College, PA 16801, USA}
\affiliation[b]{University of Wuppertal, Department of Physics, Wuppertal D-42119, Germany}
\affiliation[c]{Inst. for Theoretical Physics, ELTE E\"otv\"os Lor\' and University, P\'azm\'any P. s\'et\'any 1/A, H-1117 Budapest, Hungary}
\affiliation[d]{J\"ulich Supercomputing Centre, Forschungszentrum J\"ulich, D-52425 J\"ulich, Germany}
\affiliation[e]{E{\"o}tv{\"o}s University, Budapest 1117, Hungary}
\affiliation[f]{Department of Physics, University of Houston, Houston, TX 77204, USA
}

\emailAdd{paolo.parotto@gmail.com}

\abstract{The equation of state of Quantum Chromodynamics (QCD) at finite density  
is currently known only in a limited range in the baryon chemical potential $\mu_B$. This 
is due to fundamental shortcomings of traditional methods such as Taylor expansion 
around $\mu_B=0$. In this contribution, we present an alternative scheme 
\cite{Borsanyi:2021sxv} that displays substantially improved convergence over the Taylor 
expansion method.  We calculate the alternative expansion coefficients in the 
continuum, and show our results for the thermodynamic observables up to 
$\mu_B/T\le3.5$.}

\FullConference{%
 The 38th International Symposium on Lattice Field Theory, LATTICE2021
  26th-30th July, 2021
  Zoom/Gather@Massachusetts Institute of Technology
}

\begin{document}
\maketitle

\section{Introduction}

The QCD equation of state has been the subject of intense effort from the lattice 
community for over a decade. At vanishing baryon density, where we know the transition 
is a smooth crossover  \cite{Aoki:2006we}, continuum extrapolated results with physical 
quark masses have been available for a few years, with results from different 
collaborations showing excellent agreement 
\cite{Borsanyi:2010cj,Borsanyi:2013bia,HotQCD:2014kol}. 
At non-zero chemical potential, though, owing to the infamous complex action problem, 
direct simulations have not been possible. Thus, results at finite baryon density have 
been produced by relying on some kind of extrapolation from simulations at zero or 
imaginary chemical potentials.
Although promising new techniques are being developed, that allow for direct 
simulations at finite chemical potential, these  have not yet been applied to 
large-scale QCD simulations \cite{Sexty:2019vqx, Giordano:2020uvk, Giordano:2020roi,Borsanyi:2021hbk}.

The primary means by which the extrapolation to finite chemical potential is carried out
are Taylor expansion and analytical continuation from imaginary chemical potential 
\cite{Allton:2002zi, Borsanyi:2012cr, deForcrand:2002hgr, DElia:2002tig}. 
In the Taylor expansion method, as the name suggests, $\hmu_B$-derivatives
\footnote{We use the following notation for the dimensionless chemical potentials: 
$\hmu_i = \mu_i / T$.} of the partition function are calculated, from which 
thermodynamic quantities can then be reconstructed 
\cite{Borsanyi:2018grb,Bazavov:2020bjn}. 
In the case of analytical continuation from imaginary chemical potential, quantities of 
interest are calculated at different values of the chemical potential itself, then an 
extrapolation is carried out to describe the behavior at finite (real) chemical potential,
often resulting in improved signal. A recent example is the study of the transition line
of QCD, which is now known to great accuracy up to next-to-leading order in the 
baryon chemical potential, for which the two methods are in good agreement, although 
the analytical continuation from imaginary chemical potential shows improved 
uncertainty \cite{HotQCD:2018pds,Borsanyi:2020fev}. 

Other approaches, beyond the lattice allow for the study of the QCD phase diagram. For 
temperatures of $T \gtrsim 300\MeV$, perturbative methods are in quantitative 
agreement with lattice results (see e.g., Refs.~\cite{Bellwied:2015lba,Haque:2020eyj}). 
Functional methods also provide an alternative for studying the QCD phase diagram 
\cite{Dupuis:2020fhh}. Especially in the transition region of QCD, though, lattice 
methods still represent the major tool of investigation.

The equation of state of QCD is of intrinsic interest in the study of the properties of 
strongly interacting matter. Furthermore, it plays an crucial role in the study of heavy-ion 
collisions, as well as in the physics of high-density objects such as neutron stars 
\cite{Dexheimer:2020zzs}. 
In heavy-ion collision experiments, the phase diagram of QCD can be probed by varying 
the collision energy. Crucially, the equation of state is a fundamental input for the 
hydrodynamic simulations used to study the experimental results, but its knowledge 
from first principles is still limited in its $\hmu_B$ range. Significant computational 
efforts was placed with the goal of extending the reach of the Taylor expansion, yet even 
the sixth $\hmu_B$-derivative of the QCD pressure is currently available with modest 
precision \cite{Borsanyi:2018grb,Bazavov:2020bjn}.  

We illustrate in these proceedings a new  scheme for extrapolating the equation of state 
of QCD to larger chemical potential, which we devised to be tailored to the specific 
problem at hand \cite{Borsanyi:2021sxv}.  We are then able to considerably improve 
the convergence properties when compared to the Taylor approach, and thus reach 
larger values of the chemical potential. We present continuum extrapolated results
for thermodynamics observables at chemical potentials as high as  $\hmu_B = 3.5$.

\section{The alternative expansion scheme}
\label{sec-1}
\begin{figure}
\begin{minipage}{0.49\linewidth}
\includegraphics[width=1.01\textwidth]{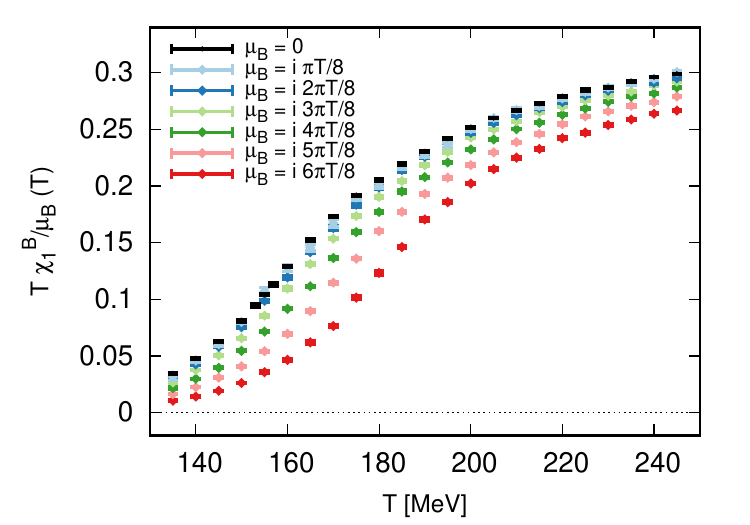} 
\end{minipage}
\begin{minipage}{0.51\linewidth}
\vspace{2.8mm}
\includegraphics[width=1.04\textwidth]{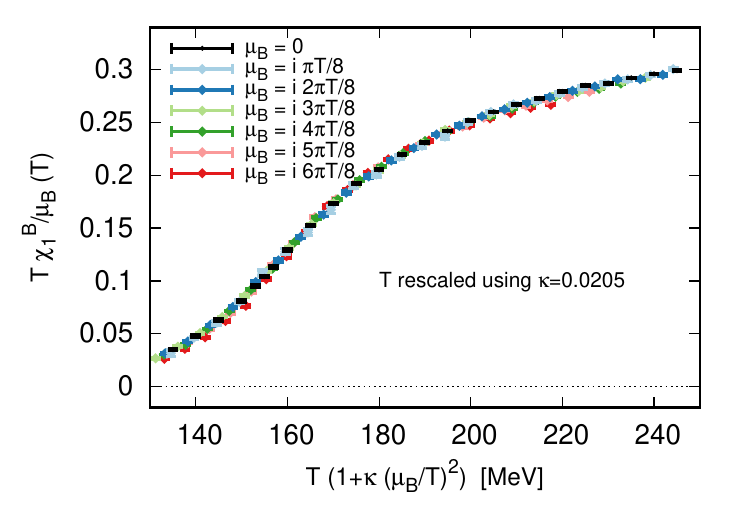} 
\end{minipage}
\begin{minipage}{0.49\linewidth}
\includegraphics[width=1.01\textwidth]{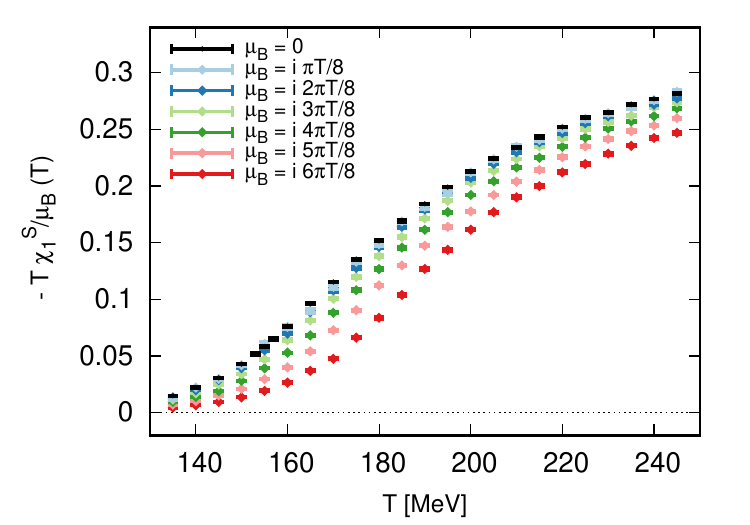} 
\end{minipage}
\begin{minipage}{0.49\linewidth}
\vspace{0.6mm}
\includegraphics[width=1.01\textwidth]{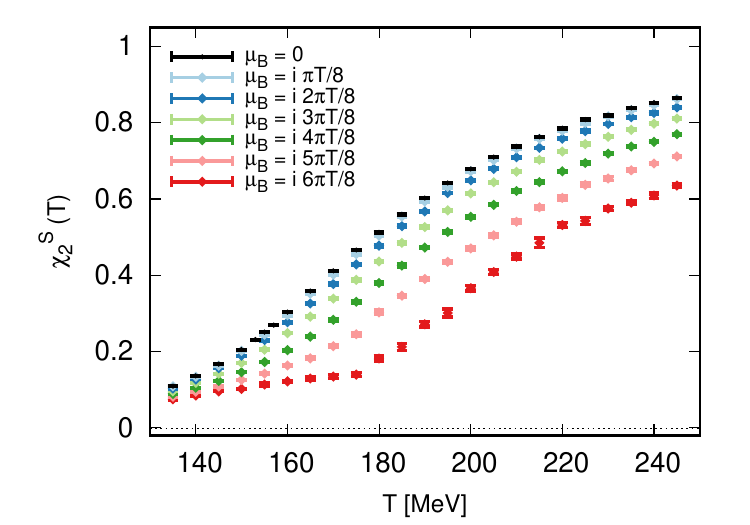} 
\end{minipage}
\vspace{-0mm}
\caption{(Top left panel): Normalized baryon density at imaginary baryon chemical 
potentials. The points at $\mu_B=0$ (black) show the second baryon susceptibility 
$\chi_2^B(T)$. (Top right panel): Same as in the left panel, with the temperature 
rescaled in accordance to Eq.~\eqref{eq:rescale1} with $\kappa=0.0205$. (Bottom left 
panel): Normalized strangeness density at imaginary baryon chemical 
potentials. The points at $\mu_B=0$ (black) show the mixed baryon strangeness 
susceptibility $\chi_{11}^{BS}(T)$. (Bottom right panel): Second strangeness 
susceptibility $\chi_{2}^{S}(T)$ at zero and imaginary baryon chemical.} 
\label{fig:final_kN}
\end{figure}

The starting point of our procedure is noticing that certain fluctuation observables 
display a similar structure at all the imaginary chemical potentials we simulate. This is the 
case for both the baryon and strangeness densities (normalized by $\hmu_B$), as well
as for the second strangeness susceptibility, as shown in Fig.~\ref{fig:final_kN}.  For 
the (normalized) baryon and strangeness densities, one can easily notice that the limit 
$\hmu_B \rightarrow 0$ equals $\chi_2^B$ and $\chi_{11}^{BS}$, respectively.

Let us take a look at the (normalized) baryond density first. As a consequence of the 
similarity of its temperature dependence at different chemical potential, its $\hmu_B$ 
behavior can be captured through a $\hmu_B$-dependent rescaling of the temperature. 
In fact, a single temperature-independent parameter is able to describe such rescaling 
quite well:
\begin{equation} \label{eq:rescale1}
\frac{\chi_1^B(T,\hmu_B)}{\hmu_B} = \chi_2^B (T^\prime, 0) \, \, , \qquad T^\prime = (1 + \kappa \, \hmu_B^2) \, \, .
\end{equation} 
Such a simple rescaling of the temperature makes all the curves collapse onto each 
other, as shown in the top right panel of Fig.~\ref{fig:final_kN}. Analogous observations 
can be made for both $\chi_1^S$ and $\chi_2^S$, with:
\begin{equation} \label{eq:rescaleSTR}
\frac{\chi_1^S(T,\hmu_B)}{\hmu_B} = \chi_{11}^{BS} (T^\prime, 0) \, \, , \qquad \chi_2^S(T,\hmu_B) = \chi_2^S (T^\prime, 0) \, \, .
\end{equation} 

The fact that a single parameter, not depending on the temperature, can describe the 
shift we observe in the observables of interest, suggests that the next-order parameter
would probably be sensibly smaller in magnitude. In order to define our expansion 
scheme in a rigorous manner, we add higher order expansion parameters, and let
them depend on the temperature:
\begin{equation} \label{eq:Tprime2}
T^\prime = T \left(1 + \kappa_2^{ij} (T) \, \hmu_B^2 + \kappa_4^{ij} (T) \, \hmu_B^4 + \calO (\hmu_B^6) \right) \, \, ,
\end{equation}
where the $k_n^{ij}(T)$ now indicate that they can refer to either one of the observables 
we consider.

We stress that the expansion scheme we propose is essentially a reorganization of the 
Taylor series. Rather than carrying out the expansion at $T=\textit{const}$ as in the 
Taylor method, we follow ``lines of constant physics'', namely trajectories in the 
$T-\hmu_B$ plane where e.g.,  $\chi_1^B(T,\hmu_B)/\hmu_B$ is constant. In fact, the 
rhs of e.g., Eq.\eqref{eq:rescale1} is effectively expanded in $\Delta T = T^\prime - T$.
Working out expansions of both the lhs and rhs, then equating equal order terms, one 
finds
\begin{equation}\label{eq:k2k4}
\kappa_2^{BB} (T) = \frac{1}{6 T} \frac{\chi_4^B (T)}{{\chi_2^B}^\prime (T)} \, \, ,  \quad\quad
\kappa_4^{BB} (T) = \frac{1}{360 \, {{\chi_2^B}^\prime (T)}^3} \left( 3 \, {{\chi_2^B}^\prime (T)}^2 \chi_6^B (T) - 5 \, {\chi_2^B}^{\prime \prime}\!\! (T) \, {\chi_4^B (T)}^2 \right) \, \, ,
\end{equation}
with analogous relations holding for the other observables.

\section{Results}
\label{sec-2}

\begin{figure}
\center
\includegraphics[width=0.33\linewidth]{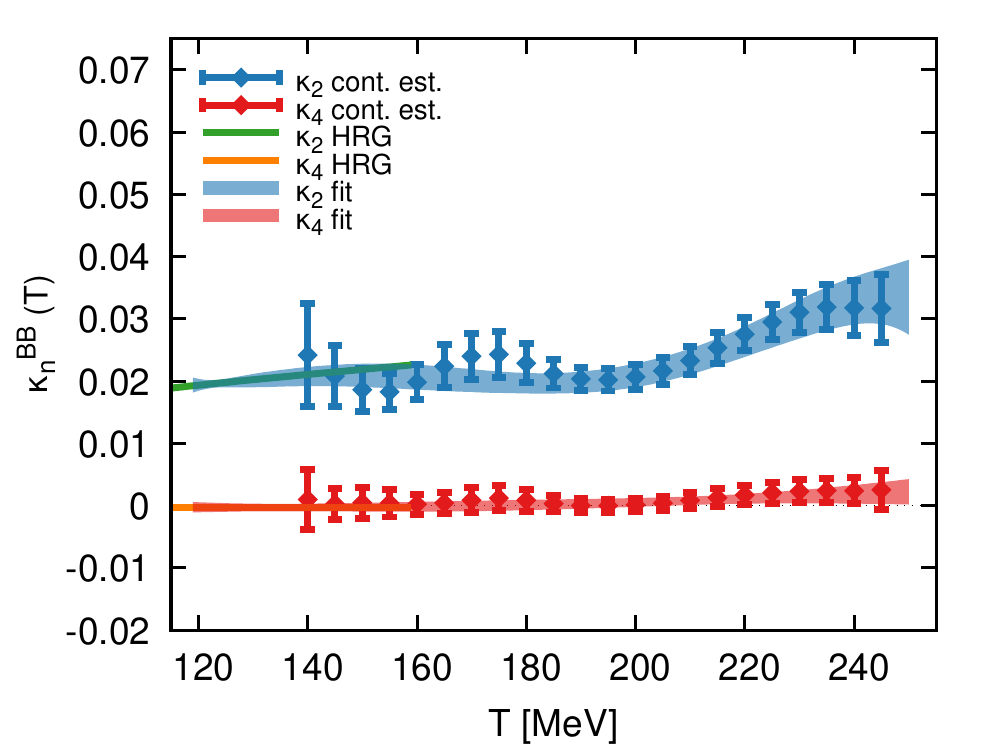}
\includegraphics[width=0.33\linewidth]{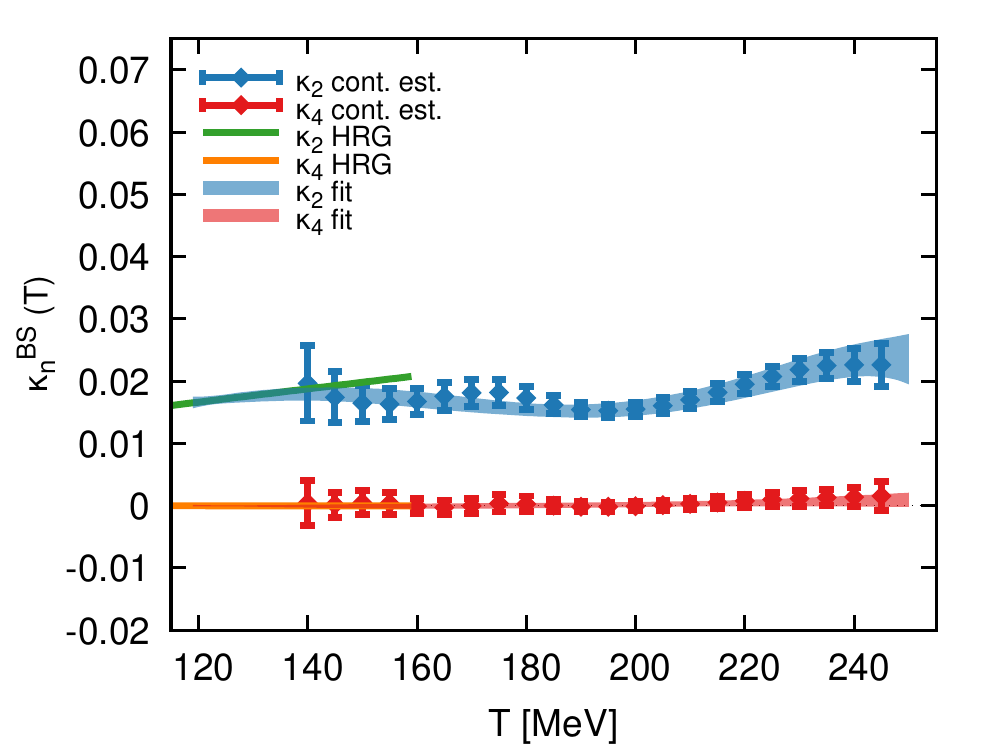}\includegraphics[width=0.33\linewidth]{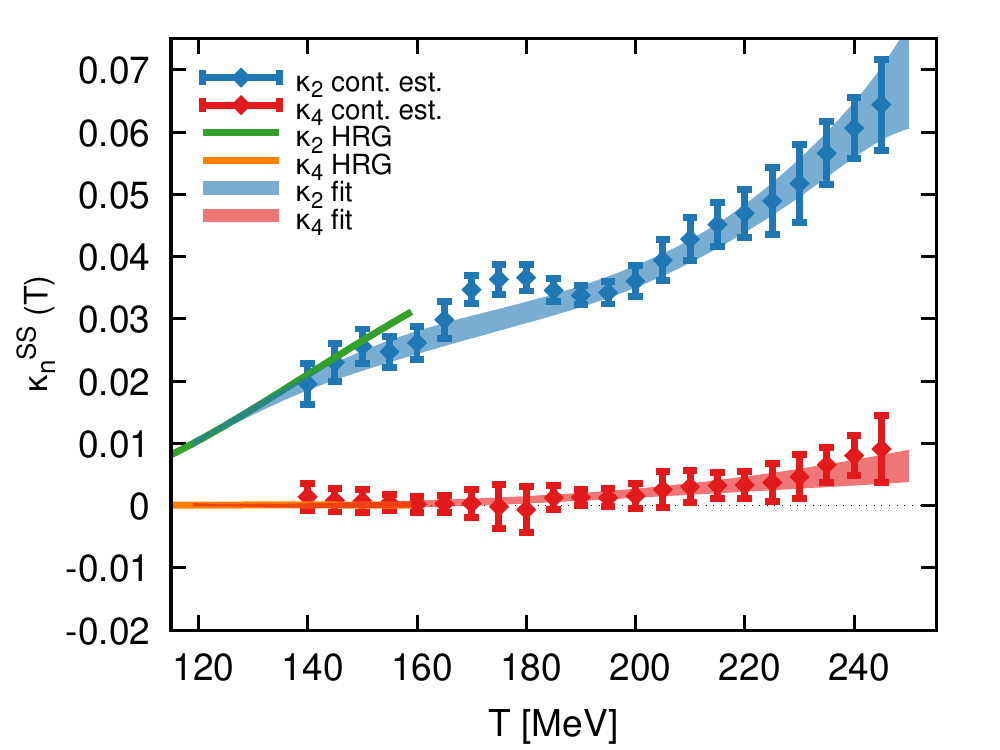}
\caption{Continuum extrapolated result for the expansion parameters 
$\kappa_2^{ij}(T)$ and $\kappa_4^{ij}(T)$. HRG results are shown up to 
$T=160\MeV$ (in green for $\kappa_2^{ij}$, orange for $\kappa_4^{ij}$,
respectively). The bands show correlated polynomial fits.}
\label{fig:kappaN}
\end{figure}

In order to calculate $\kappa_n^{ij}(T)$, we could simply take Eq.\eqref{eq:k2k4} at face 
value. However, large cancellations would likely appear in $\kappa_4^{ij}(T)$. Instead, we 
choose to exploit our simulations at imaginary chemical potentials 
$\hmu_B = i \, n \, \pi/8$, with $n = 0,...,8$. First, we construct the proxy quantity:
\begin{equation}
\Pi (T, \hmu_B^2) = \frac{T^\prime-T}{T\hmu_B^2} = \kappa_2 (T) + \kappa_4 (T) \hmu_B^2  + \calO (\hmu_B^4) \, \, ,
\end{equation}
at different temperatures and chemical potentials. At $\hmu_B=0$, we simply have 
$\Pi (T, 0) = \kappa_2^{ij}(T)$. At finite $\hmu_B$, $T^\prime$ is first determined by 
solving e.g., Eq.~\eqref{eq:rescale1} for $T^\prime$. For this, we first perform a spline fit 
to the data at zero and finite chemical potentials. 
Then, we can calculate $\Pi (T, \hmu_B^2)$ for different chemical potentials, 
temperature-by-temperature. We include results from $32^3\times8$, $40^3\times10$, 
$48^3\times12$ and $64^3\times16$ lattices.

Finally, we perform a combined fit of $\Pi (T, \hmu_B^2)$ in $\hmu_B^2$ and 
$1/N_\tau^2$, from which we extract the continuum-extrapolated $\kappa_2^{ij}$ and 
$\kappa_4^{ij}$ at each temperature. 
In order to perform a systematic analysis of the errors, we carry out our procedures 
many times, then combine the results to include both statistical and systematic 
uncertainties.  We include three different spline fits at $\hmu_B=0$, and two at 
$\hmu_B \neq 0$. For the scale setting, we utilize either $f_\pi$, or $w_0$
\cite{Borsanyi:2012zs}. As ranges in 
the imaginary chemical potential included in the fit, we consider
$\mathrm{Im}~\hmu_B\le 2.0$ or $\mathrm{Im}~\hmu_B\le 2.4$. 
For the final fit, we consider fits that are linear in $1/N_\tau^2$, and linear, quadratic or 
1/linear for the continuum limit. Finally, we either include or drop the $N_\tau=8$ 
results. Considering all possible combinations, this amounts to 144 separate analyses. 
After dropping those with a Q-values below 0.01, we weigh the results uniformly to 
obtain our final results. For a more detailed account of the treatment of the systematics, 
see Refs.~\cite{Borsanyi:2018grb, Borsanyi:2021sxv}.

The different coefficients $\kappa_n^{ij}(T)$ are shown in Fig.~\ref{fig:kappaN}, 
together with the results of polynomial fits which take into account the full correlations
between different temperatures. We show at low temperature results from the 
hadron resonance gas (HRG) model, in good agreement with our results. 
As expected, we find a large separation in the magnitudes of $\kappa_2^{ij}(T)$ and 
$\kappa_4^{ij} (T)$, suggesting that convergence should be faster with this expansion.

\begin{figure*}[!]
\includegraphics[width=0.49\linewidth]{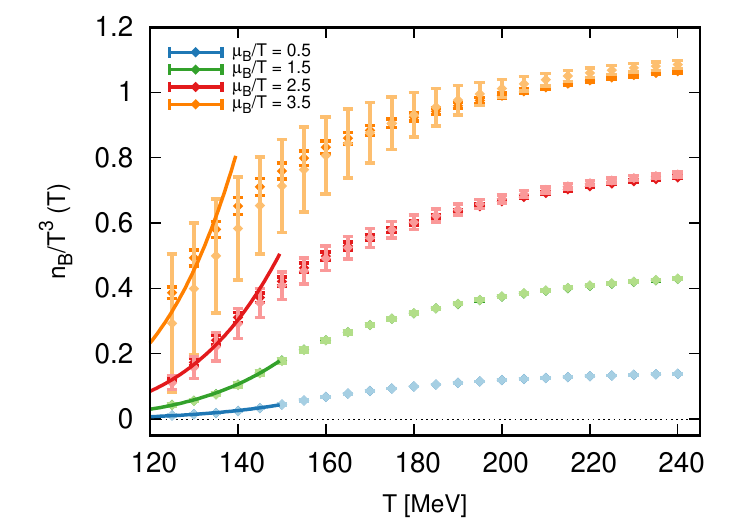}
\includegraphics[width=0.49\linewidth]{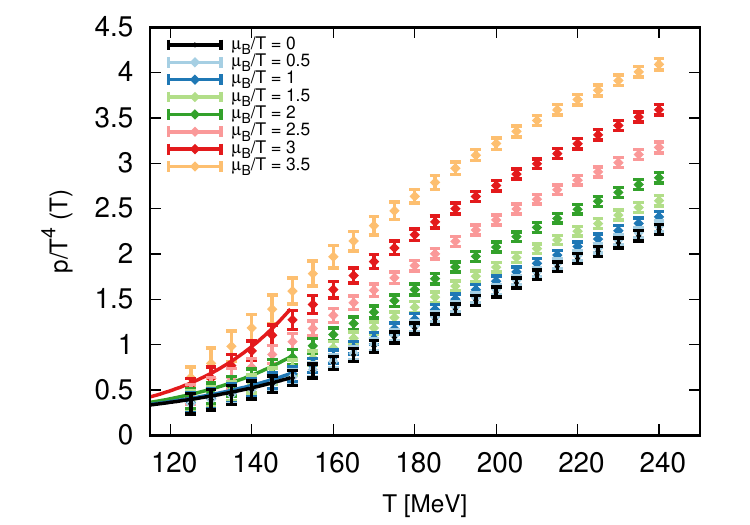}
\includegraphics[width=0.49\linewidth]{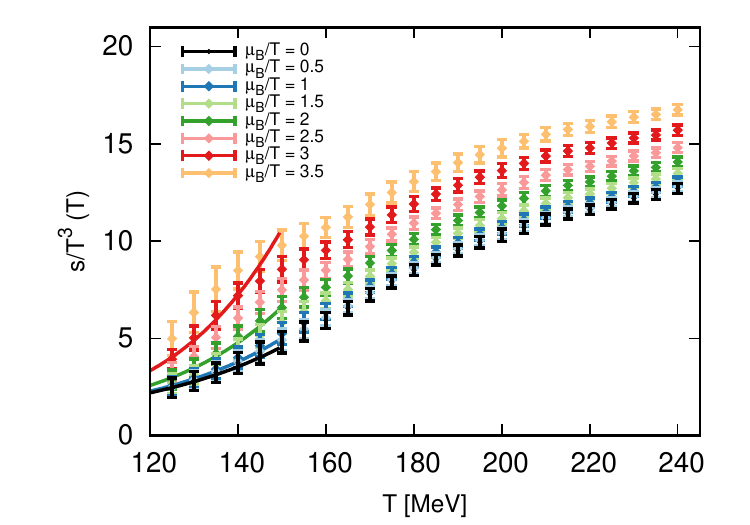}
\includegraphics[width=0.49\linewidth]{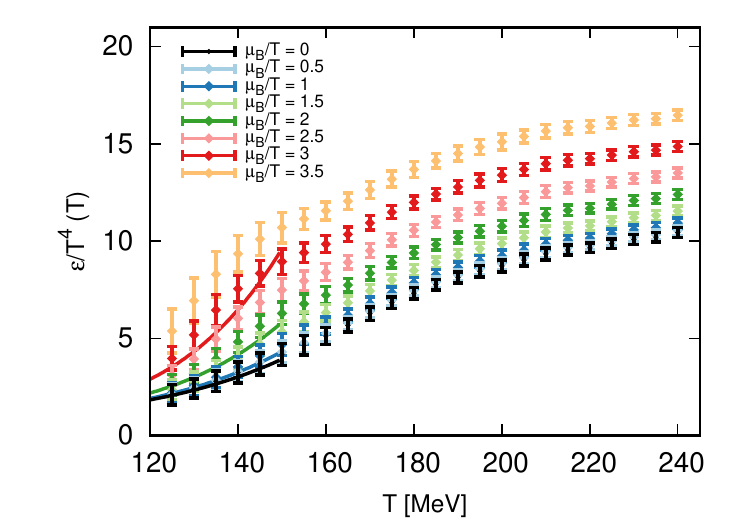}
\includegraphics[width=0.49\linewidth]{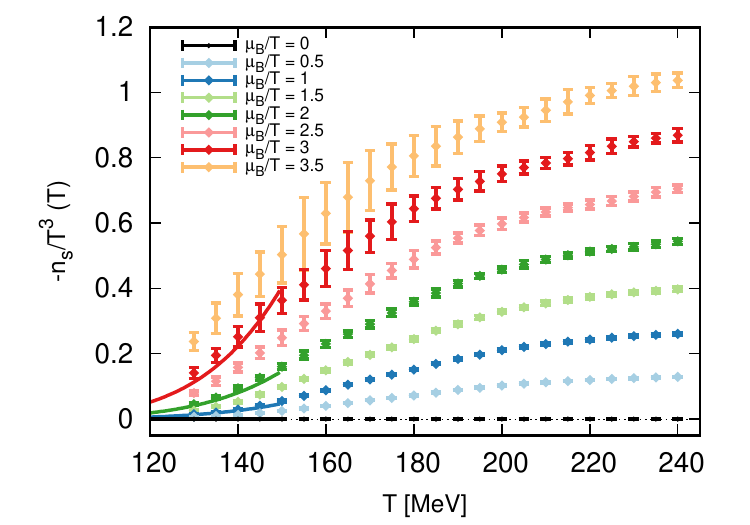}
\includegraphics[width=0.49\linewidth]{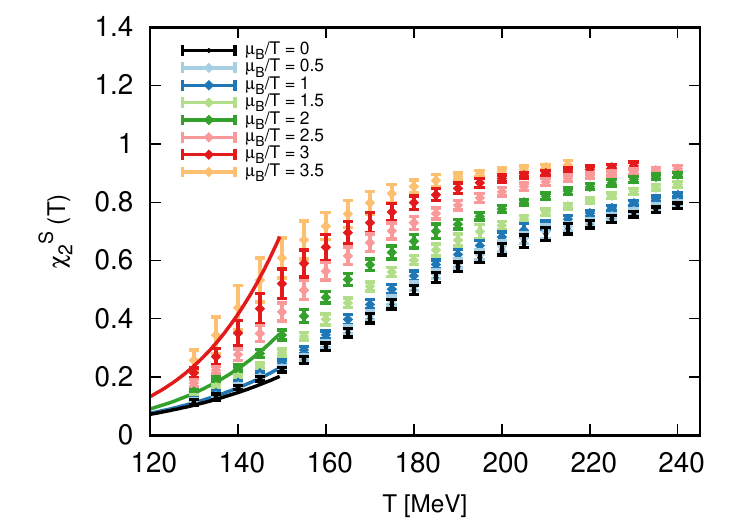}
\caption{From top to bottom, left to right: baryon density, pressure, entropy density, 
energy density, strangeness density and strangeness second susceptibility. The results
are shown at increasing values of $\hmu_B$. 
With solid lines we show the results from the HRG model. For the baryon density we also 
show in darker shades the results obtained by setting $\kappa_4^{BB} \equiv 0$.}
\label{fig:thermo_mux}
\end{figure*}

To reconstruct thermodynamic quantities at real chemical potential, we simply
employ Eq.~\eqref{eq:rescale1} (similarly, for the strangeness density and second 
susceptibility, we employ Eq~\eqref{eq:rescaleSTR}), starting from $\chi_2^B(T)$. We 
then obtain the pressure by integrating over the chemical potential:
\begin{equation}
\frac{p(\mu_B,T)}{T^4} = \hat{p}(\hmu_B,T) = \hat{p}(0,T) + \bigintssss_0^{\hmu_B} \!\!\!\!\!\! {\rm d} \hmu^\prime_B \, \hat{n}_B (\hmu_B^\prime,T) \, \, ,
\end{equation}
and all other thermodynamic quantities follow.

Fig.~\ref{fig:thermo_mux} shows the baryon density (top left), pressure (top right), entropy density (center left), energy density (center right), strangeness density (bottom left) and strangeness second susceptibility (bottom right) in the range 
$T = 130-240 \MeV$, for $\hmu_B \leq 3.5$. We see that the uncertainties are under 
control in all cases, and good agreement with the HRG model is observed also at finite 
$\hmu_B$. Moreover, no unphysical, nonmonotonic behavior is present. For the baryon
density, we additionally show (darker shades) the case in which $\kappa_4^{BB}=0$. We
notice that the effect of including this additional parameter is only that of increasing the
error, while the value itself is barely affected. This further suggests that our series 
expansion possesses good convergence properties.

\section*{Acknowledgments}
This project was funded by the DFG grant SFB/TR55.
The project also received support from the BMBF Grant No. 05P18PXFCA.
This work was also supported by
the Hungarian National Research,  Development and Innovation Office, NKFIH
grant KKP126769. A.P. is supported by the J. Bolyai Research
Scholarship of the Hungarian Academy of Sciences and by the \'UNKP-20-5 New
National Excellence Program of the Ministry for Innovation and Technology.
The project leading to this publication has received funding from
Excellence Initiative of
Aix-Marseille University - A*MIDEX, a French ``Investissements d'Avenir''
programme, AMX-18-ACE-005.
This  material is based upon  work  supported  by  the National  Science
Foundation under grants no. PHY-1654219 and by the U.S. DoE, 
Office  of  Science,  Office  of  Nuclear  Physics, within the framework of the
Beam Energy Scan Topical (BEST) Collaboration.  
This research used resources of the Oak Ridge Leadership Computing Facility, which is a DOE Office of Science User Facility supported under Contract DE-AC05-00OR22725. 
The authors gratefully
acknowledge the Gauss Centre for Supercomputing e.V.  (www.gauss-centre.eu) for
funding this project by providing computing time on the GCS Supercomputer
HAWK at HLRS, Stuttgart. 
Part of the computation was performed on the QPACE3 funded by the DFG ind
hosted by JSC.
C.R. also acknowledges the support from the Center of Advanced
Computing and Data Systems at the University of Houston.

%
%
%

\end{document}